\documentclass[%draft % uncomment if draft mode is needed
               ]{jetp} % jetp.cls
\twocolumn % comment if two-column layout not needed
\begin{document}
\English

\title{Peculiarities of Laue Diffraction of  Neutrons in Strongly Absorbing Crystals}

% Authors
\author{A.~Ya.}{Dzyublik}
\email{dzyublik@ukr.net}
\affiliation{Institute for Nuclear Research, NAS of Ukraine, \\
03680, prospect Nauki 47, Kiev, Ukraine}

\author{V.~V.}{Mykhaylovskyy}
\affiliation{Institute for Nuclear Research, NAS of Ukraine, \\
03680, prospect Nauki 47, Kiev, Ukraine}

\author{V.~Yu.}{Spivak}
\affiliation{Institute for Nuclear Research, NAS of Ukraine, \\
03680, prospect Nauki 47, Kiev, Ukraine}

%\keywords{neutron optics, neutron resonance, Laue diffraction, dynamical scattering theory, spherical waves, Borrmann triangle, anomalous transmission}

\abstract{%
Well-known Kato's theory of the Laue diffraction of spherical x-ray waves is generalized to the case of the neutron diffraction in strongly absorbing crystals, taking into consideration both the potential  and the resonant scattering of neutrons by nuclei as well as a realistic angular dispersion of incident neutrons.
The saddle-point method is applied for estimation of the angular integrals, being more adequate in the case of strongly absorbing crystals than the usually used stationary-phase approximation. It is found that the intensity distribution  of the diffracted and refracted beams  along the basis of the Borrmann triangle significantly depends on the deviation of the neutron energy from the  nuclear resonant level. When comparing our calculations with the Shull's experimental data on  neutron diffraction in silicon we regard also the role of  finite width of the collimating and scanning slits. }

\maketitle   % please do not remove

\section{Introduction}

The neutron scattering is one of the most powerful tools for investigation of the crystal structure and its magnetic properties (see, e.g., Refs.~[1-6]). For interpretation of experimental data, obtained in very thin films, it is sufficient to apply the kinematical scattering theory, while for thick targets, where the multiple scattering of neutrons by atoms becomes significant, one has to use already its dynamical version. Such a dynamical scattering theory has been developed for the elastic diffraction in perfect crystals of both neutrons [7-9] and M\"{o}ssbauer rays [10-13], treating  them as plane waves. Its generalization to the case of inelastic diffraction at the crystals subject to external alternating fields was given in  Refs.~\cite{Dz1,Dz2}.

In the case of two-wave diffraction  it was predicted the  effect of suppression of reactions  and inelastic channels \cite{Kagan,Kagan1},  observed later in numerous M\"{o}ssbauer diffraction experiments (see, e.g., the reviews \cite{Belyakov,Rudolf}) as well as
     in the  neutron diffraction experiments \cite{exp1,exp2}, studying the $(n,\gamma)$ reaction  in the cadmium sulphide crystal,
  abandoned  with the nuclei $^{113}$Cd having the resonant level  0.178 eV.

   Although there is close analogy of the suppression of nuclear reactions with the  anomalous transmission of x-rays (the Borrmann effect) [20-25],
      the resonant nuclear scattering provides principally new character of the anomalous absorption of neutrons or $\gamma$-photons. Namely,
   in the scattering of x-rays by atomic electrons, when the Bragg condition is fulfilled,  there is only partial suppression of inelastic scattering.  At the same time,  in the case of resonant nuclear scattering, choosing the corresponding geometry of the experiment, one can  achieve complete suppression of the inelastic and reaction  channels.
   This effect is explained by the dynamical scattering theory, which predicts that the  energy exchange between the refracted and diffracted waves inside the crystal ensures splitting each  of them  into  two waves with different wave vectors. One such wave is weakly absorbed and another strongly.

Notice once more that in Refs.~[7-15]  the electromagnetic waves and neutrons were described by plane waves.
At the same time, in typical experiments on the  Laue diffraction  the incident waves first pass a narrow slit and
only afterwards penetrate into the crystal (see Fig.1). In this case the incident neutrons
 are described already by a wave packet written as the integral over the angle. Moreover, both the refracted and diffracted waves  travel
 inside the crystal within the region, confined by  so-called Borrmann triangle [20-25].
  The distribution of diffracted beam intensity along the basis of the Borrmann triangle oscillates due to
   interference of two waves, transmitting  the crystal with different wave vectors. The same interference provides
  also the familiar Pendell\"{o}sung effect [20-24].

   Kato [26-29] developed a theory for such a diffraction, assuming  the
   angular dispersion of incident x-ray waves to be much larger than a small diffraction interval of the order of several seconds of arc. Although Kato told about the diffraction of spherical waves emitted by a point-like source,  in reality he considered the cylindrical waves, emitted by the thread-like source. More exactly, the collimating entrance slit has been regarded like the continuous line of such point emitters, stretched along the slit.

 Shull  [30-32] used the Kato's theory  in order to interpret the results of his  Laue diffraction experiments of neutrons
  in perfect crystals of silicon  and germanium. Measuring the period of interference oscillations for the diffracted neutron beam he determined
  with high precision the coherent scattering lengths of neutrons by the nuclei of silicon  \cite{Shull1} and germanium \cite{Shull3}.
  These experiments were repeated later by Abov and Elyutin \cite{Abov,Abov1}.

  Worth noting also the papers [33-37], studying  the Laue diffraction of neutrons in large silicon crystals at the Bragg angle close to $\pi/2$.
  In these conditions the authors observed anomalous absorption and slowing down of neutrons.
 Moreover, they hoped to verify the equivalence principle of the inertial and gravitational
masses on the example of such microscopic object as  neutron and achieve higher accuracy $\sim 10^{-5}$.

  Previously we  analyzed the symmetric Laue diffraction of divergent neutron beams \cite{DSM}  in strongly absorbing crystals,
    taking into account both the potential and  resonant  neutron scattering by nuclei.  Following Kato [26-29], we took the angular dispersion $\sigma_a$ of the incident neutrons much exceeding the characteristic angular interval $|\Delta\vartheta|$, where the  diffraction proceeds. In other words, a real angular distribution of neutrons in this approximation was replaced by a constant.
     The analogous theory for the symmetric Laue diffraction of the M\"{o}ssbauer radiation has been developed in \cite{DS}.

  In the present paper we study general case of the Laue diffraction, when the reflecting crystal planes are canted at arbitrary angle with respect to the crystal surface. For the first time  the angular distribution of incident neutrons $G_a(\theta)$  is included into consideration with dispersion $\sigma_a$, which may be of the order of the diffraction angular range $|\Delta\vartheta|$.
    For calculation of the angular integrals, which determine the neutron wave packet inside  a thick crystal, we use the saddle-point method. It is more adequate for  strongly absorbing crystals than the  stationary-phase method, which has been  used previously in  the  theory of  x-rays diffraction in weakly absorbing crystals \cite{Authier}. Note  that in the Kato's approximation, when the function $G_a(\theta)$ is replaced by a constant, these angular integrals are calculated exactly \cite{Authier}. However, in our approach due to the factor $G_a(\theta)$ we need   estimations of the integrals by means of the saddle-point approximation for thick enough crystals.

    We consider typical experimental situation (see, e.g., [30-32]), when the neutrons pass first through the entrance slit cut in a shield and then move inside the crystal within the Borrmann triangle. We assume that all the neutrons fall perpendicularly to the collimating slit, which in turn is parallel to the reflecting crystal planes. First we shall   regard the slit as a thread-like emitter.
    In this case the neutron waves have a cylindrical symmetry with the symmetry axis $z$ along the slit.
    And in the fourth section we analyze the role of finite width of both slits used in experiments.

    \section{Basic formulas}
Let the incident neutrons at $t \to -\infty$ be described by the initial wave packet
% [h]=here, [t]=top of page, [b]=bottom of page.
% width=\linewidth,height=\linewidth
\begin{figure}
\includegraphics[width=\linewidth]{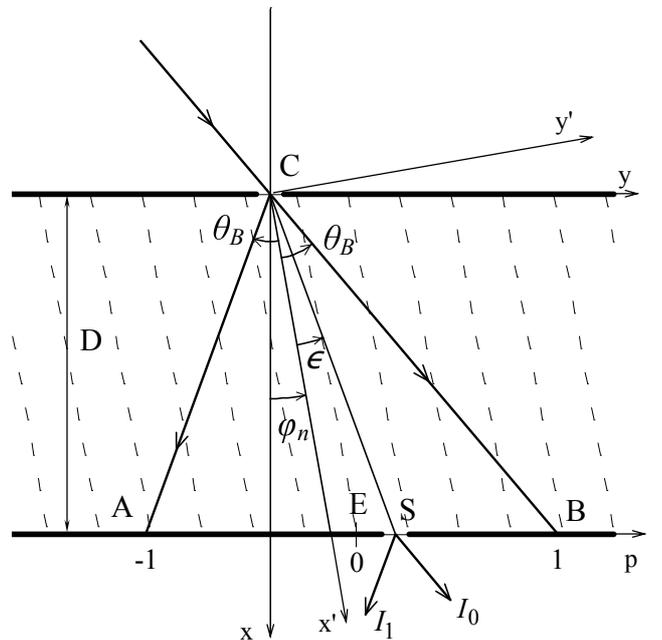}
\caption{Scheme of the Laue diffraction of the collimated neutron wave in a perfect crystal.The flow of neutrons is concentrated mainly inside  the Borrmann triangle ABC. The collimating and scanning slits are labeled by C and S, respectively.The reflecting planes are drawn by the dashed lines. The points A and B are marked by the reduced coordinates $p=-1$ and $p=1$, respectively; the middle point E of the line segment AB by $p=0$.}
\end{figure}

\begin{equation}\label{eq:Psi}
\Psi_{{\textrm{in}}}({\bf r},t)
=\int \frac{d{\boldsymbol\kappa}}{(2\pi)^3}
f({\boldsymbol\kappa})e^{i{\boldsymbol\kappa}{\bf r}-iE t/\hbar},
\end{equation}
where ${\boldsymbol\kappa}$ are the wave vectors of neutrons,  $E=\hbar^2\kappa^2/2m$ is their energy and $m$  the  mass. The product $|f({\boldsymbol\kappa})|^2\Delta{\boldsymbol\kappa}$ is interpreted as a probability of finding the neutron with the wave vector ${\boldsymbol\kappa}$ lying in the interval $\Delta{\boldsymbol\kappa}$ beside ${\boldsymbol\kappa}$.
 For brevity, we omit the spin factor, which does not  affect the coherent scattering by nonmagnetic crystal with unpolarized nuclei.

We introduce the  right-hand coordinate frame $x,\;y,\;z$  with the origin on the entrance
 surface  in the middle of the collimating slit. The axis $x$ is directed inside the crystal perpendicularly to its
 surface and the axis $z$  along the slit (see Fig.1).
 One introduces also the frame $x',\;y',\;z'$  with the axis $x'$  parallel to  the reflecting crystal planes and axis $z'$ coinciding with $z$. It is obtained from $x,\;y,\;z$ by  rotation through the angle $\varphi_n$ around the axis $z$. Here $\varphi_n<0$ for clockwise rotation  and $\varphi_n>0$ otherwise. The angle between the neutron wave vector ${\boldsymbol\kappa}$  and the axis $x$ is $\varphi$, the  incidence angle on the reflecting planes is $\theta$ and the Bragg angle $\theta_{\textrm{\scriptsize {B}}}.$

  Besides, we introduce the angles $\varphi_0$  and $\varphi_1$  between the axis $x$ and the sides of the Borrmann triangle, corresponding to transmitting and diffracted rays ($\varphi_0=\theta_{{\scriptsize {B}}}+\varphi_n >0$ and
 $\varphi_1=-\theta_{\textrm{\scriptsize {B}}}+\varphi_n<0 $).

 Let the divergent beam of neutrons move in the plane $x,~y$ perpendicularly to the slit and be spread over the angle $\theta$. Then $f({\boldsymbol \kappa })\sim \delta(\kappa_z)$,
 while  the components of ${\boldsymbol\kappa}(\theta)$ along  the axes $x,\;y,\;z$  are
   \begin{equation}
{\boldsymbol \kappa }(\theta)=\{\kappa\cos\varphi,\; \kappa\sin\varphi,\; 0\},
  \end{equation}
where $\varphi=\varphi_n+\theta$.

The wave function (\ref{eq:Psi})  may be rewritten as
\begin{equation}\label{eq:WP1}
\Psi_{{\textrm{in}}}({\bf r},t)=\int_0^{\infty} G_{e}(E)\Psi^{{\textrm{in}}}_E({\bf r})e^{-iE t/\hbar}dE,
\end{equation}
where $G_{e}(E)$ characterizes the energy  distribution of
incident neutrons, the function
\begin{equation}\label{eq:WP2}
\Psi^{\textrm{in}}_E({\bf r})=\int_{-\pi}^{\pi}
G_a(\theta)e^{i{\boldsymbol \kappa }(\theta){\bf r}} d\theta
\end{equation}
  describes neutrons with fixed energy $E$.
 We approximate  the angular distribution by the Gaussian function with maximum at the angle $\theta_0$
  close to  the Bragg angle $\theta_{\textrm{\scriptsize {B}}}$:
\begin{equation}\label{eq:Gauss}
G_a(\theta)=\frac{1}{(\sqrt{2\pi}\sigma)^{1/2}}\exp\left\{-\frac{(\theta_{0}-\theta)^2}{4\sigma^2}\right\},
\end{equation}
where
\begin{equation}
\sigma^2=\langle (\theta_0-\theta)^2 \rangle
\end{equation}
 denotes the mean-square angular dispersion of neutrons.
Usually $\sigma<<1$, that enables us to spread the integration limits over $\theta$ from ${-\infty}$ to $\infty$.
If $\sigma$ much exceeds the  diffraction  interval $|\Delta\vartheta|$, then  $G_a(\theta)$ can be replaced by a constant.

 The energy distribution is  also  described by the Gaussian function:
\begin{equation}
G_{e}(E)= \frac{1}{(\sqrt{2\pi}\sigma_e)^{1/2}}\exp\left\{-\frac{(E-\bar{E})^2}{4\sigma_e^2}\right\}.
\end{equation}

We shall consider scattering of the neutron wave  by  atomic nuclei in the crystal, ignoring influence of the electrons.
Then the coherent scattering of neutrons by an elementary cell of the crystal from the state  ${\boldsymbol\kappa}$ to ${\boldsymbol\kappa}'$  is determined by
the amplitude
\begin{equation}\label{eq11}
F({\boldsymbol\kappa},{\boldsymbol\kappa}')=\sum_je^{i{\bf Q}{\boldsymbol\rho}_j}{\bar f}_{j}({\boldsymbol\kappa},{\boldsymbol\kappa}'),
\end{equation}
where ${\bf Q}={\boldsymbol\kappa}-{\boldsymbol\kappa}'$ is the scattering wave vector,
the  radius-vector ${\boldsymbol\rho}_j$  defines equilibrium position of the $j$th atom within the elementary cell,
${\bar f}_{j}({\boldsymbol\kappa},{\boldsymbol\kappa}')$ is the coherent scattering amplitude of low-energy neutrons by the $j$th nucleus:
\begin{equation}
{\bar f}_{j}({\boldsymbol\kappa},{\boldsymbol\kappa}')=-{\bar b}_je^{-W_j({\bf Q})}
+{\bar f}^{{\textrm{res}}}_{j}({\boldsymbol\kappa},{\boldsymbol\kappa}'),
\end{equation}
where ${\bar b}_j$ is the coherent scattering length of neutrons by the $j$th nucleus,
$e^{-W_j({\bf Q})}$ is the square root of the Debye-Waller factor, ${\bar f}_{j}^{{\textrm{res}}}({\boldsymbol\kappa},{\boldsymbol\kappa}')$ is
the coherent resonant scattering amplitude. In  vicinity of an isolated resonance it is given by
\begin{eqnarray}\label{eq:fj}
\qquad{\bar f}_{j}^{{\textrm{res}}}({\boldsymbol\kappa},{\boldsymbol\kappa}')=
-c_j\left(\frac{2I_e+1}{2I_g+1}\right)\frac{\Gamma_n}{2\kappa_0}\qquad\qquad
\end{eqnarray}
$$
\times \sum_{\{n'_s\}}
\left\langle\frac{(\exp[-i{\boldsymbol\kappa}'{\bf u}_j])_{\{n^0_s\}\{n'_s\}}
(\exp[i{\boldsymbol\kappa}{\bf u}_j])_{\{n'_s\}\{n^0_s\}}}
{E-E_0-\sum_s\hbar\omega_s(n'_s-n^0_s)+i\frac{\Gamma}{2}} \right \rangle,\nonumber
$$
where  $c_j$ is the probability of finding the resonant isotope in the $j$th site, $I_g$ is the spin of the ground state  of the initial nucleus and $I_e$ the  spin of the excited  state of the compound nucleus, $E_0=\hbar^2\kappa_0^2/2m$ is  the energy  of the resonant level,  $\Gamma=\Gamma_n+\Gamma_{\gamma}+\Gamma_f$  is the total width, consisting of the partial widths for  neutron
$\Gamma_n$, radiation $\Gamma_{\gamma}$ and possibly fission $\Gamma_f$ decay channels, ${\bf u}_j$ is the displacement of the $j$th nucleus from the equilibrium position,
$\{n_s^0\}$ and $\{n'_s\}$ are sets of phonon numbers in the initial and final states of the crystal, $\omega_s$ are the phonon frequencies,
the brackets $\langle...\rangle$ denote averaging over the initial states of the crystal lattice.

The sum in Eq.(\ref{eq:fj}) can be transformed to the integral:
\begin{eqnarray}\label{eq:S1}
\sum_{\{n'_s\}}\left\langle...\right\rangle=
-ie^{-W_j({\boldsymbol\kappa})}e^{-W_j({\boldsymbol\kappa}')}\\
\times\int_0^{\infty}\frac{dt}{\hbar}e^{i(E-E_0)t/\hbar-\Gamma t/2\hbar+\varphi_j(t)},\nonumber
\end{eqnarray}
where $e^{-W_j({\boldsymbol\kappa})}$ is the Lamb-M\"{o}ssbauer factor,
\begin{multline}\label{eq:v1}
\varphi_j(t)=\sum_s\frac{\hbar}{2M_jN\omega_s}\times \\ \times
\left[y_{js} {\bar n}_s e^{i\omega_st}+
y_{js}^*({\bar n}_s+1)e^{-i\omega_st}
 \right]
\end{multline}
and
\begin{equation}
y_{js}=({\boldsymbol\kappa}{\bf v}_{js})({\boldsymbol\kappa}'{\bf v}^*_{js}),
\end{equation}
 $M_j$ is the mass of the $j$th atom, $N$ is the number of elementary cells, ${\bar n}_s$ is the average number of phonons of the $s$th normal vibration with polarization ${\bf v}_{js}$.

 In the framework of the Debye model of the crystal with one atom per the elementary cell, ignoring  anisotropy of vibrations, one can rewrite this expression as (see also \cite{A})
 \begin{eqnarray}
 \varphi_j(t)=\frac{3}{2}\frac{({\bf p}\cdot{\bf p}')}
 {M (k_B\Theta_{\textrm{D}})^3}\int_0^{\omega_{\textrm{max}}}\hbar^2\omega  d\omega\\
 \times \left[{\bar n}_{\omega}e^{i\omega t}+({\bar n}_{\omega}+1)e^{-i\omega t}
 \right], \nonumber
 \end{eqnarray}
where ${\bf p}=\hbar{\boldsymbol\kappa}$ and ${\bf p}'=\hbar{\boldsymbol\kappa}'$ are the initial and final momenta of neutrons, $\Theta_{\textrm{D}}$  is the Debye temperature, $\omega_{\textrm{max}}=k_B\Theta_{\textrm{D}}/\hbar$ is the maximal frequency of phonons.

In such  Debye approach  the  parameter $W(\kappa)=(1/2)\kappa^2\langle u_x^2\rangle $ is given by \cite{Zaiman}
\begin{align}\label{eq:Deb}
W(\kappa)=
\frac{3\cal{R}}{k_B\Theta_D}\left(\frac{T}{\Theta_D}
  \right)^2
  \int_0^{\Theta_{\textrm{D}}/T}\left[\frac{1}{e^z-1}+\frac{1}{2}\right]zdz,
\end{align}
where   ${\cal R}=\hbar^2\kappa^2/2M$ represents the recoil energy of the nucleus with  mass $M$.

In the approximation of fast collisions, when $\hbar\omega_{\textrm{max}}/\Gamma$ $<<1$, the expression (\ref{eq:S1}) reduces to
\begin{equation}\label{eq:ap}
\sum_{\{n'_s\}}\left\langle...\right\rangle=\frac{e^{-W_j({\bf Q})}}{E-E_0+i\frac{\Gamma}{2}}.
\end{equation}
This approximation is well fulfilled for low-lying resonances. Specifically, for the $^{113}$CdS crystal with parameters of the neutron resonance $E_0=0.1779 \pm 0.0002$~eV, $\Gamma_n=0.638 \pm 0.0008$~meV, $\Gamma_{\gamma}=112.4\pm 0.4$~meV \cite{Cd}  and $\Theta_{\textrm{D}}=219$K  one has $\hbar\omega_{\textrm{max}}/\Gamma\approx 0.2$.

 \section{The   wave function}
According to collision theory \cite{Goldberger} every plane wave  $e^{i{\boldsymbol \kappa (\theta)}{\bf r}} $
   of the wave packet (\ref{eq:WP2}) is scattered independently of each other, giving rise to   the  wave function ${\boldsymbol \psi}_{{\boldsymbol \kappa }(\theta)}({\bf r})$.
Therefore the neutron wave  packet, born  by the incident wave  Eq.~(\ref{eq:WP2}) takes  the form
\begin{equation}\label{eq:120}
 \Psi_E({\bf r})=
\int_{-\infty}^{\infty}d\theta G_a(\theta)
{\boldsymbol \psi}_{{\boldsymbol \kappa} (\theta)}({\bf r}),
\end{equation}

In the two-wave case  the wave  ${\boldsymbol \psi}_{{\boldsymbol \kappa }(\theta)}({\bf r})$ inside the crystal  as $0<x<D$, where $D$ is the crystal thickness, consists of the refracted wave with the wave vector
 ${\bf k}(\theta)$ and the diffracted one with the wave vector ${\bf k}_{1}(\theta)={\bf k}(\theta)+{\bf h}$,  where ${\bf h}$ denotes a reciprocal lattice vector. The components of the vectors ${\bf k}(\theta)$ and ${\boldsymbol\kappa}(\theta)$
along the entrance surface $x=0$ coincide. Therefore the vectors ${\bf k}_{\nu}(\theta)$ with $\nu=0,1$  can be written as
\begin{equation}\label{eq:}
{\bf k}_{\nu}(\theta)={\boldsymbol\kappa}_{\nu}(\theta) +\delta(\theta){\bf n},\qquad
{\boldsymbol\kappa}_{1}(\theta)={\boldsymbol\kappa}_{0}(\theta)+{\bf h},
\end{equation}
where ${\bf n}$ is the unit vector along the axis $x$.

As a consequence,  the wave function $ \Psi_{E }({\bf r})$ inside the crystal transforms to
\begin{eqnarray}\label{eq:Psi1}
\Psi_E({\bf r})=
\sum_{\nu=0,1}\Psi_{E}^{(\nu)}({\bf r}),\nonumber\qquad\qquad \\
\Psi_E^{(\nu)}({\bf r})=\int_{-\infty}^{\infty} d\theta G_a(\theta)
\psi_{{\boldsymbol \kappa}_{\nu}(\theta)}({\bf r}),
\end{eqnarray}
with $\psi_{{\boldsymbol \kappa}_{\nu}(\theta)}({\bf r})$ given by
\begin{align}\label{eq:66}
 \psi_{{\boldsymbol \kappa}_{\nu}(\theta)}({\bf r})=
 \sum_{\iota=1,2}C_{\nu}^{(\iota)}(\theta)
e^{i{\boldsymbol\kappa}_{\nu}(\theta){\bf r}+i\delta_{\iota }(\theta)x}.
\end{align}

Following \cite{Kagan} we introduce  the notations
\begin{equation}
 k_0(\theta)=\kappa[1+\varepsilon_0(\theta)], \qquad k_1(\theta)=\kappa[1+\varepsilon_1(\theta)].
\end{equation}
The parameters $\varepsilon_{0(1)}(\theta)$   are related by
\begin{equation}\label{eq:12}
\varepsilon_1=\alpha/2+\gamma_1\varepsilon_0/\gamma_0,
\end{equation}
where
\begin{equation}
\alpha=\frac{2{\boldsymbol\kappa}(\theta){\bf h}+{\bf h}^2}{\kappa^2},
\qquad \gamma_{\nu}=\cos\varphi_{\nu}.
\end{equation}
The $1+\varepsilon_0(\theta)$ means the refractive index for the incident wave $\exp\{i{\boldsymbol \kappa}_{0}(\theta){\bf r}\}$.

Recall that $\varphi_{\nu}$ are the angles between the vectors ${\boldsymbol\kappa}_{\nu}={\boldsymbol\kappa}_{\nu}(\theta_{\textrm{\scriptsize {B}}})$ and the axis $x$.
The angle $\alpha$ indicates  deviation from the exact Bragg condition corresponding to $\kappa_1=\kappa$.
For neutrons with fixed energy \cite{Zach}
\begin{equation}\label{eq:Z}
\alpha=2\sin2\theta_B\Delta\theta,
\end{equation}
where
\begin{equation}\label{eq:1d}
 \Delta\theta=\theta_{\textrm{\scriptsize {B}}}-\theta.
\end{equation}

The corrections to the wave numbers in the medium  $\delta$ are related with the parameters $\varepsilon_0$  by
\begin{equation}\label{eq:11}
\delta(\theta)= \kappa\varepsilon_{0}(\theta)/\gamma_0.
\end{equation}

The amplitudes $C$ and the wave vectors ${\bf k}$ are determined by the system of fundamental equations of the dynamical scattering theory \cite{Authier}.
For the two-wave case in notations of Ref.~\cite{Kagan} they are written as
\begin{eqnarray}\label{eq:100}
 (k^2(\theta)/\kappa^2(\theta)-1)C_{0}=g_{00}C_{0}+g_{01}C_{1}, \nonumber \\
(k^2_1(\theta)/\kappa^2(\theta)-1)C_{1}=g_{10}C_{0}+g_{11}C_{1}.
\end{eqnarray}
The scattering   matrix $g_{\mu\nu}$ is defined by the expression
\begin{equation}\label{eq:3}
g_{\mu\nu}=\frac{4\pi }{\kappa^2
v_0}F({{\boldsymbol\kappa}_{\nu},{\boldsymbol\kappa}_{\mu}}),\qquad\qquad
\mu,\nu=0,1,
\end{equation}
where $v_0$  stands for the volume of the elementary cell.

The system of two  equations (\ref{eq:100}) has the
following solution  \cite{Kagan}:
\begin{eqnarray}\label{eq:23}
\varepsilon_{0}^{(1,2)}=\frac{1}{4}\left[g_{00}+
\beta g_{11}-\beta\alpha \right]
\pm \frac{1}{4}\Large\{\left[g_{00}+\beta g_{11}
\right. \\ \left.
-\beta\alpha\right]^2
+4\beta\left[g_{00}\alpha
-(g_{00}g_{11}-g_{01}g_{10})\right]\Large\}^{1/2},\nonumber
\end{eqnarray}
where
\begin{equation}
 \beta=\gamma_0/\gamma_1.
\end{equation}
Henceforth the root with sign plus is associated with $\varepsilon_{0}^{(1)}$ and minus with $\varepsilon_{0}^{(2)}$.

It is more convenient to  express them in terms of new deviation parameter
\begin{equation}\label{eq:4}
\eta=\frac{1}{2}\left(\frac{\beta}{g_{01}g_{10}}\right)^{1/2}(\alpha-\alpha_0),
 \end{equation}
 where the angular shift
 \begin{equation}
 \alpha_0=g_{11}  -g_{00}/\beta.
 \end{equation}
 Note that $\eta$ is already a complex number.

Then the parameters $\varepsilon_{0}^{(1,2)}$ take simple form
\begin{equation}
\varepsilon_{0}^{(\iota)}=\frac{1}{2}g_{00}-\frac{1}{2}\sqrt{g_{01}g_{10}\beta}
\left[\eta+(-1)^{\iota+1}\sqrt{1+\eta^2}   \right]
\end{equation}
and  $\delta_{\iota}(\eta)$, defined by Eq.(\ref{eq:11}), may be written as
  \begin{equation}\label{eq:delta}
\delta_{\iota}(\eta)=\frac{\kappa g_{00}}{2\gamma_0}-\frac{\pi}{\Lambda_L}\left[\eta+(-1)^{\iota+1}\sqrt{1+\eta^2}\right],
\end{equation}
where
\begin{equation}
\Lambda_L=\frac{2\pi \gamma_0}{\kappa \sqrt{ g_{01}g_{10}\beta}}
\end{equation}
means the Pendell\"{o}sung distance in the case of weakly
absorbing crystals  (see, e.g.,
\cite{Authier}).

For the Laue diffraction ($\beta>0$) the amplitudes of the waves satisfy the following boundary condition at $x=0$:
\begin{equation}
\sum_{\iota=1,2}C_{0}^{(\iota)}(\theta)=1, \qquad \sum_{\iota=1,2}C_{1}^{(\iota)}(\theta)=0.
\end{equation}
Being expressed in terms of the variable $\eta$, they  take the form
\begin{align}\label{eq:C}
C_{0}^{(\iota)}(\eta) &=\frac{1}{2}\left[1+(-1)^{\iota}\frac{\eta}{\sqrt{1+\eta^2}}\right], \nonumber
\\
C_{1}^{(\iota)}(\eta) &=\frac{(-1)^{\iota}}{2}\left(\frac{g_{10}}{g_{01}}\right)^{\frac{1}{2}}\sqrt{\frac{\beta}{1+\eta^2}}.
\end{align}

Let us express the angular distribution $G_a(\theta)$ as a function  of  $\eta$.
By means of Eqs. (\ref{eq:Z}) and (\ref{eq:4}), one finds first the relation between the departures $\Delta\theta$ and $\eta$:
\begin{equation}\label{eq:333}
\theta_{\textrm{B}}-\theta=\Delta\vartheta\eta-
\Delta\theta_{\textrm{B}},
\end{equation}
where
\begin{equation}
\Delta\vartheta=\frac{1}{\sin 2\theta_{\textrm{B}}}
\sqrt{\frac{g_{01}g_{10}}{\beta}}, \qquad \Delta\theta_{\textrm{B}}=-\frac{\alpha_0}{2\sin 2\theta_{\textrm{\scriptsize {B}}}}.
\end{equation}
Here $|\Delta\vartheta|$
has a sense of the  characteristic diffraction range, corresponding to variation of $|\eta|$ from zero to unity.

According to (\ref{eq:C}) the maximal amplitude of the diffracted
neutron wave is achieved at $\eta=0$, i.e. if
\begin{equation}\label{eq:334}
\theta=\theta'_{\textrm{B}}, \qquad\qquad \theta'_{\textrm{B}}=
\theta_{\textrm{B}}+\Delta\theta_{\textrm{B}}.
\end{equation}
From here we see that  $\theta'_{\textrm {B}}$ can be interpreted
as  the Bragg angle shifted by $\Delta\theta_{\textrm{B}}$.

Assuming that the incident beam is oriented along the corrected
Bragg angle,  $\theta_0=\theta'_{\textrm{B}}$, we rewrite the
angular distribution as
\begin{equation}
G_a(\theta) \to {\cal G}_a(\eta)=
 \frac{1}{(2\pi\overline{\eta^2})^{1/4}}\exp\left\{-\frac{\eta^2}{4\overline{\eta^2}}\right\},
\end{equation}
where the mean-square width
\begin{equation}
\overline{\eta^2}=\left(\sigma/\Delta\vartheta\right)^2.
\end{equation}
From definition of $\eta$ it follows also that
\begin{equation}
G_a(\theta)d\theta= -\sqrt{\Delta\vartheta}{\cal G}_a(\eta)d\eta.
\end{equation}

The neutron intensity distribution over the basis of the Borrmann
triangle is usually analyzed with the aid of the scanning slit,
located on the rear surface and  directed  along the axis z. Let ${\bf r}_p=\{D,\;y_S,\;z\}$
be the radius vector of any point $S$
inside this slit, while $y_0$ be the coordinate of the
midpoint E on the  side AB of the Borrmann triangle. Following  \cite{Authier}  we determine the reduced coordinate of
 this point as
\begin{equation}\label{eq:p11}
p=\frac{\Delta y_S}{L},
\end{equation}
where $2L$ is the length of the line segment AB, $\Delta y_S=y_S-y_0$ is the coordinate of the point $S$
relative to the midpoint E. The definition (\ref{eq:p11}) is
equivalent to
\begin{equation}\label{eq:p}
p=2\frac{\Delta y_S/D}{\tan\varphi_0-\tan\varphi_1},
\end{equation}
which  in the case of symmetric diffraction, $\beta=1$,  reduces to  the
definition of $p$, given in Refs.~\cite{Batterman,Kato1}:
\begin{equation}
p=\frac{\tan\epsilon}{\tan\theta_{\textrm{\scriptsize {B}}}},
\end{equation}
where  $\epsilon$ is the angle between the reflecting planes and
the direct line CS,  connecting  the entrance slit and the point $S$
(see Fig.~1).

It remains now to expand the plane wave $e^{i{\boldsymbol\kappa}(\theta){\bf r}}$ in the point ${\bf r} \approx {\bf r}_S$
in  powers of  $\Delta\theta$.  Keeping the linear terms in the expansion of ${\boldsymbol\kappa}(\theta){\bf r}$
and introducing the notation ${\boldsymbol\kappa}_{\nu}={\boldsymbol\kappa}_{\nu}(\theta_B)$
we get
\begin{eqnarray}\label{eq:p10}
e^{i{\boldsymbol\kappa}(\theta){\bf r}}=
\exp\Large\left[i\kappa D\sin\varphi_0\left(1-\frac{y_s/D}{\tan\varphi_0} \right)\Delta\theta \right]\\
\times \exp\{i{\boldsymbol\kappa}_0{\bf r}\}.\qquad\qquad\qquad\qquad\quad\nonumber
\end{eqnarray}

With the help of relations
\begin{eqnarray}
\tan\varphi_0+\tan\varphi_1=\frac{\sin 2\varphi_n}{\gamma_0\gamma_1}
\end{eqnarray}
and
\begin{eqnarray}
\tan\varphi_0-\tan\varphi_1=\frac{\sin 2\theta_B}{\gamma_0\gamma_1},
\end{eqnarray}
which follow from  the equalities
\begin{equation}
\varphi_0+\varphi_1=2\varphi_n, \qquad \varphi_0-\varphi_1=2\theta_B,
\end{equation}
we find that
\begin{equation}\label{eq:xz}
\Delta y_S/D=\frac{1}{2\gamma_0\gamma_1}[\sin 2\varphi_n+p\sin 2\theta_B ].
\end{equation}

Substitution of this formula into  (\ref{eq:p10}) after some trigonometric manipulations gives
\begin{eqnarray}\label{eq:pw}
\exp\{{i{\boldsymbol\kappa}(\theta){\bf r}}\} =
\exp\left\{i\frac{\kappa D}{4\gamma_1}(1-p)\alpha_0\right\}\nonumber \\ \times
\exp\left\{i\frac{\pi D}{\Lambda_L}(1-p)\eta\right\}e^{i{\boldsymbol\kappa}_0{\bf r}}.\qquad\qquad
\end{eqnarray}

Taking also into account Eq.~(\ref{eq:delta}), we are led to the following expression for the waves in the exit slit:
\begin{eqnarray}
\exp\{{i{\boldsymbol\kappa}(\theta){\bf r}+i\delta_{\iota}(\theta)D}\} = \Phi(p;E)\qquad\qquad\qquad \\ \times
\exp\left\{-i\frac{\pi D}{\Lambda_L}\left[p\eta+(-1)^{\iota+1}\sqrt{1+\eta^2}\right] \right\}e^{i{\boldsymbol\kappa}_0{\bf r}},\nonumber
\end{eqnarray}
where we used the abbreviation
\begin{equation}
\Phi(p;E)=\exp\left\{i\frac{\kappa D}{4}\left[\frac{g_{00}} {\gamma_0}
+\frac{g_{11}}{\gamma_1}
 +p\left(\frac{g_{00}} {\gamma_0}-\frac{g_{11}} {\gamma_1} \right)\right]\right\}.
 \end{equation}

Substituting  (\ref{eq:pw}) into Eqs.~(\ref{eq:Psi1}),  (\ref{eq:66}) and introducing
 the  notations
\begin{equation}
{\cal N}=\frac{\pi D}{|\Lambda_L|},
\end{equation}
 \begin{equation}
{\cal S}_{\iota}(\eta)=-i\left(|\Lambda_L|/\Lambda_L\right) \left[p\eta+
(-1)^{\iota+1}\sqrt{1+\eta^2}\right],
\end{equation}
one finds the integral representation for the wave function in the
point ${\bf r}\approx {\bf r}_S$:
\begin{eqnarray}\label{eq:10}
\Psi_E^{(\nu)}({\bf r})=-\Phi(p;E)\sqrt{\Delta\vartheta}
\end{eqnarray}
$$ \times
 \int_{C}d\eta{\cal G}_a(\eta)
\sum_{\iota=1,2}C_{\nu}^{(\iota)}(\eta) e^{{\cal N}
{\cal S}_{\iota}(\eta)}e^{i{\boldsymbol\kappa}_{\nu}{\bf
r}},\qquad\qquad\qquad
$$
 where the integration path $C$ in the complex plane $\eta=\eta_r+i\eta_i$ is a direct line,  defined by the condition Im~$[\alpha_0+2(g_{01}g_{10}/\beta)^{1/2}\eta]=0.$

For a  crystal, whose thickness $D>>|\Lambda_L|/\pi$, the  large parameter ${\cal N}$  allows us to
estimate the integral over $\eta$ with the aid of the saddle-point method
(see, e.g., \cite{Math}). Here we assume that $C(\eta)$ as well as $G_a(\theta)$ are smooth functions.
 First from the equation ${\cal S}'_{\iota}(\eta)=0$
 one finds the saddle points:
\begin{equation}
\eta^{(\iota)}_0=(-1)^{\iota}\frac{p}{\sqrt{1-p^2}}.
\end{equation}
Since the integrand in (\ref{eq:10}) is an analytical function one can deform the integration contour $C$ on  the complex plane
$\eta$. This contour should cross
the $\iota$th saddle point along the line which indicates a steepest decent of the
function ${\cal S}_{\iota}(\eta)$. Along such a line Im~${\cal S}_{\iota}(\eta)=$const and the function Re~${\cal S}_{\iota}(\eta)$ is maximal in the point $\eta_0$.
 These requirements are satisfied if the line is directed with respect to the
real axis $\eta_r$ at the angle
\begin{eqnarray}\label{eq:Var}
\vartheta_{\iota}=\pm \frac{\pi}{2}
-\frac{1}{2}\mbox{arg}~{\cal S}''_{\iota}(\eta_0),
\end{eqnarray}
where the second derivative of ${\cal S}_{\iota}(\eta)$ in the saddle point equals
\begin{equation}\label{eq:S}
{\cal S}''_{\iota}(\eta_0)=i(-1)^{\iota}\left(\frac{|\Lambda_L|}{\Lambda_L}\right)(1-p^2)^{3/2}.
\end{equation}

Inserting (\ref{eq:S}) into (\ref{eq:Var})
one finds that
 at $|p|<1$
\begin{eqnarray}\label{eq:angle}
\vartheta_{\iota}=(-1)^{\iota}\frac{\pi}{4}+\mbox{arg}\sqrt{\Lambda_L}.
\end{eqnarray}

 Evaluating the integral (\ref{eq:10})
 with the aid of  the saddle-point method, one has
\begin{eqnarray}\label{eq:98}
 \Psi_E^{(\nu)}({\bf r})=-\sqrt{\Delta\vartheta}{\cal G}_a(\eta_0)\Phi(p;E)\qquad\qquad\qquad\qquad
 \nonumber \\ \times\sum_{\iota=1,2}
 C_{\nu}^{(\iota)}
 e^{{\cal N}
{\cal S}_{\iota}(\eta_0)}\sqrt{\frac{2\pi}{{\cal
N}|{\cal S}''_{\iota}(\eta_0)|}}
e^{i\vartheta_{\iota}}e^{i{\boldsymbol\kappa}_{\nu}{\bf r}},
\end{eqnarray}
where  the amplitudes $C_{\nu}^{(\iota)} =C_{\nu}^{(\iota)}(\eta^{\iota}_0)$ in the saddle points are
\begin{eqnarray}\label{eq:90}
C_{0}^{(\iota)}=\frac{1}{2}\left( 1+p\right), \\
C_{1}^{(\iota)}=\frac{(-1)^{\iota}}{2}\left(\frac{g_{10}}{g_{01}}\beta\right)^{\frac{1}{2}}\sqrt{1-p^2},\nonumber
\end{eqnarray}
while the angular factor
\begin{equation}
{\cal G}_a(\eta_0) =\frac{1}{(2\pi\overline{\eta^2})^{1/4}}
 \exp\left\{-\frac{1}{4\overline{\eta^2}}\frac{p^2}{|1-p^2|} \right\}.
\end{equation}

Substituting (\ref{eq:S}), (\ref{eq:angle}) and (\ref{eq:90}) into
(\ref{eq:98}), one gets  the wave function of neutrons in the point ${\bf r} \approx {\bf r}_S $ inside the
scanning slit. For the refracted neutrons the wave function is
\begin{eqnarray}\label{eq:F1}
\Psi_E^{(0)}({\bf r})= \frac{1}{2}\frac{{\cal
A}_0(p)}{(1-p^2)^{1/4}} \left(
\frac{1+p}{1-p}\right)^{\frac{1}{2}}\Phi(p;E)
\end{eqnarray}
\begin{eqnarray*}
 \times
\sqrt{\frac{2\Lambda_L}{D}}\left[e^{i\zeta(p)}+e^{-i\zeta(p)}
\right]
 e^{i{\boldsymbol \kappa}_0{\bf r}}
\end{eqnarray*}
and  for the diffracted  those
\begin{eqnarray}\label{eq:F2}
\Psi_E^{(1)}({\bf r})= \frac{1}{2}\frac{{\cal
A}_1(p)}{(1-p^2)^{1/4}}\Phi(p;E)
\end{eqnarray}
\begin{eqnarray*}
 \times
\sqrt{\frac{2\Lambda_L}{D}} \left[e^{i\zeta(p)}-e^{-i\zeta(p)}\right]
 e^{i{\boldsymbol \kappa}_1{\bf r}},\nonumber
\end{eqnarray*}
where
\begin{equation}\label{eq:z}
 \varsigma(p)=\frac{\pi D}{\Lambda_L}\sqrt{1-p^2}+\frac{\pi}{4}.
\end{equation}
and the amplitudes are
\begin{eqnarray}
{\cal A}_0(p)=-\sqrt{\Delta\vartheta}{\cal
G}_a(\eta_0),\;\;\;\;
{\cal A}_{1}(p)=\left(
\frac{g_{10}}{g_{01}}\beta\right)^{\frac{1}{2}}{\cal A}_0(p).
\end{eqnarray}

The corresponding intensities of the monochromatic neutron beams  are determined by
\begin{equation}
I_{\nu}(p;E)=|\Psi^{(\nu)}_E(p)|^2.
\end{equation}

Introducing the notation
\begin{equation}
\frac{1}{\Lambda_L}=\frac{1}{\tau_L}+i\frac{1}{\sigma_L},
\end{equation}
we get the following intensity distribution of the refracted beam  through the basis of the Borrmann triangle $(|p|<1)$:
\begin{eqnarray}\label{eq:F11}
I_0(p;E)= \frac{|{\cal
A}_0(p)|^2}{\sqrt{1-p^2}}\left(\frac{1+p}{1-p}\right)e^{-\mu D}
\frac{2|\Lambda_L|}{D}
\end{eqnarray}
$$ \times
 \left[\sinh^2\left(\frac{\pi D}{\sigma_L}\sqrt{1-p^2} \right)+\cos^2\left(\frac{ \pi D}{\tau_L}\sqrt{1-p^2}+\frac{\pi}{4}
 \right) \right],\nonumber
$$
while for the diffracted beam distribution one has
\begin{eqnarray}\label{eq:F12}
I_1(p;E)= \frac{|{\cal A}_1(p)|^2}{\sqrt{1-p^2}}e^{-\mu D}
\frac{2|\Lambda_L|}{D}
\end{eqnarray}
$$ \times
\left[\sinh^2\left(\frac{\pi D}{\sigma_L}\sqrt{1-p^2} \right)+\sin^2\left(\frac{ \pi D}{\tau_L}\sqrt{1-p^2}+\frac{\pi}{4}
 \right) \right].\nonumber
$$
Here
\begin{equation}
\mu=\frac{1}{2}\left[\frac{\mu_0}{\gamma_0}+\frac{\mu_1}{\gamma_1}+
p\left(\frac{\mu_0}{\gamma_0}-\frac{\mu_1}{\gamma_1}\right)
\right],
\end{equation}
$\mu_0$ and $\mu_1$ are the absorption coefficients
for  neutrons incident  far from the Bragg condition $(1>>|\Delta\theta|>>|\Delta\vartheta)|$, no diffraction), but having the wave vectors $\approx{\boldsymbol \kappa}_{0}$ and $\approx{\boldsymbol \kappa}_{1}$, respectively. They are determined by
\begin{equation}
\mu_{\nu}=\kappa\mbox{Im}g_{\nu\nu}=\sigma_t({\boldsymbol \kappa}_{\nu})/v_0,
\end{equation}
where $\sigma_t({\boldsymbol \kappa}_{\nu})$ is the total cross section
for scattering and absorption  of neutrons  by an elementary cell, which have in the initial state the wave vector ${\boldsymbol \kappa}_{\nu}$. In accordance with the optical theorem \cite{Sitenko}
\begin{equation}
\sigma_t({\boldsymbol \kappa}_{\nu})=\frac{4\pi}{\kappa}\mbox{Im}F({\boldsymbol \kappa}_{\nu},{\boldsymbol \kappa}_{\nu}).
\end{equation}

Outside the Borrmann triangle in close vicinity to the points $p=\pm 1$  these intensities are
\begin{eqnarray}\label{eq:F14}
I_0(p;E)= \frac{|{\cal
A}_0(p)|^2}{\sqrt{p^2-1}}\left|\frac{1+p}{1-p}\right|e^{-\mu D}
\frac{2|\Lambda_L|}{D}
\end{eqnarray}
$$ \times
 \left[\sinh^2\left(\frac{\pi D}{\tau_L}\sqrt{p^2-1} \right)+\sin^2\left(\frac{ \pi D}{\sigma_L}\sqrt{p^2-1}+\frac{\pi}{4}
 \right) \right],\nonumber
$$
and
\begin{eqnarray}\label{eq:F15}
I_1(p;E)= \frac{|{\cal A}_1(p)|^2}{\sqrt{p^2-1}}e^{-\mu D}
\frac{2|\Lambda_L|}{D}
\end{eqnarray}
$$ \times
\left[\sinh^2\left(\frac{\pi D}{\tau_L}\sqrt{p^2-1} \right)+\cos^2\left(\frac{ \pi D}{\sigma_L}\sqrt{p^2-1}+\frac{\pi}{4}
 \right) \right],\nonumber
$$

Outside the Borrmann triangle the intensity of the diffracted
neutrons (\ref{eq:F15}) in strongly
absorbing crystal  first quickly falls down and then   begins to grow with increasing
 $|p|$  due to the hyperbolic sine. But in this region our
approach is not valid, since the departure  $|\Delta\theta|$ becomes too large.

\begin{figure}[t]
\includegraphics[width=\linewidth]{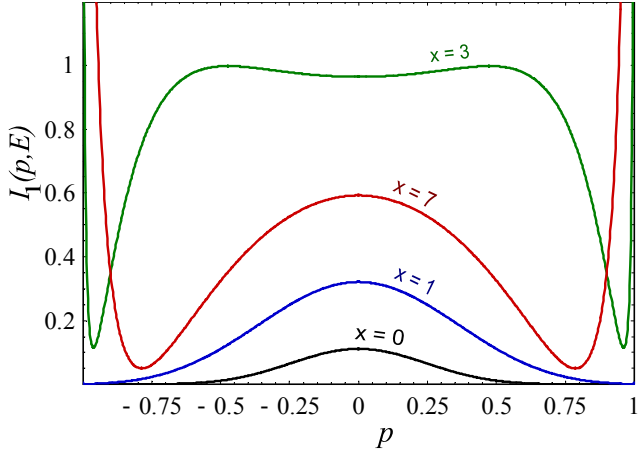}
\caption{The intensity distribution $I_1(p;E)$ of the monochromatic diffracted beam
 over the basis of the Borrmann triangle for a few values of the resonance detuning parameter $x$ at
$\mu_{\textrm{res}} D = 20$.}
\end{figure}
\begin{figure}[t]
\includegraphics[width=\linewidth]{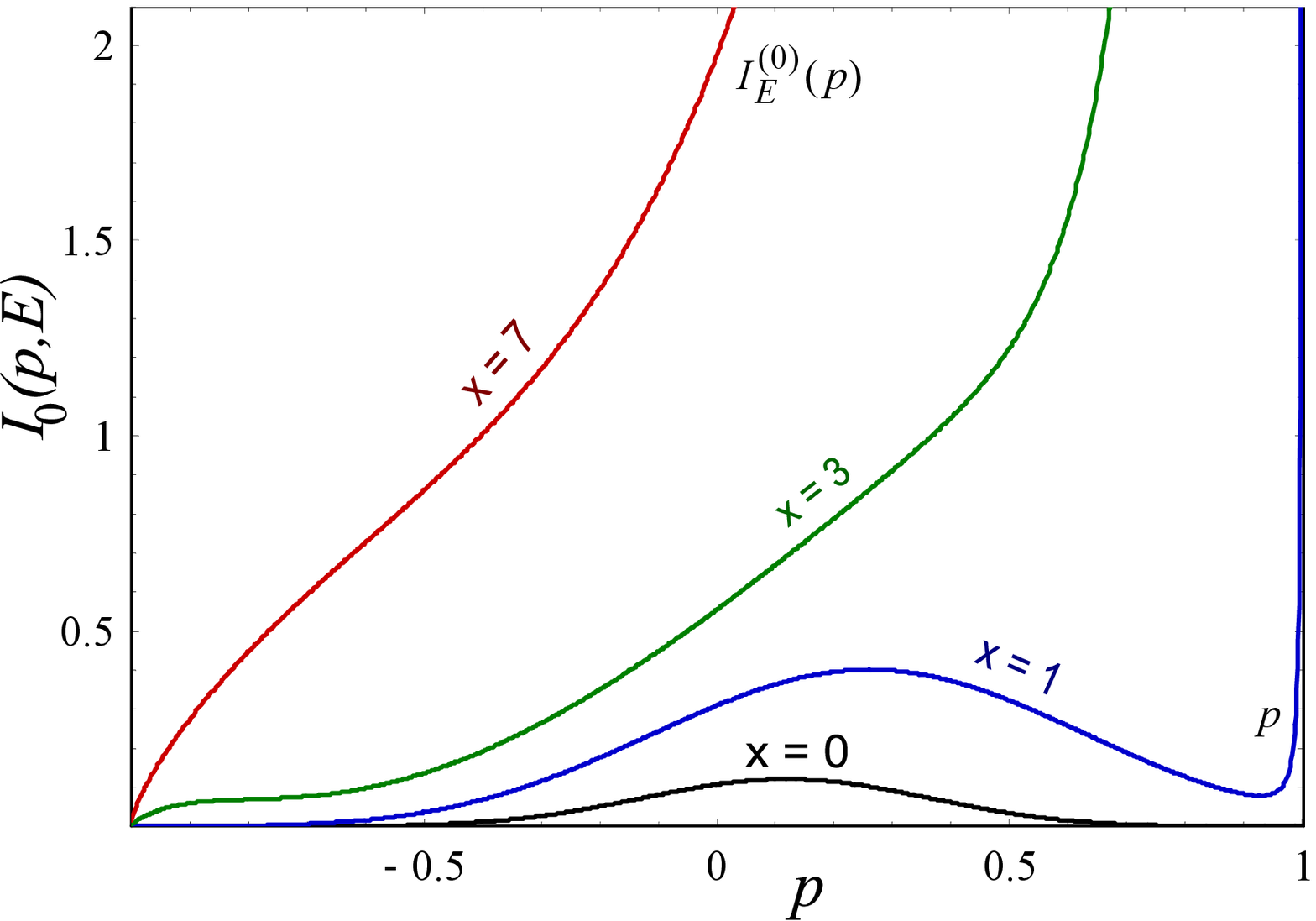}
\caption{The intensity distribution $I_0(p;E)$ of the monochromatic refracted beam
 over the basis of the Borrmann triangle for a few values of the resonance detuning parameter $x$ at
$\mu_{\textrm{res}} D = 20$. }
\end{figure}
\begin{figure}[t]
\includegraphics[width=\linewidth]{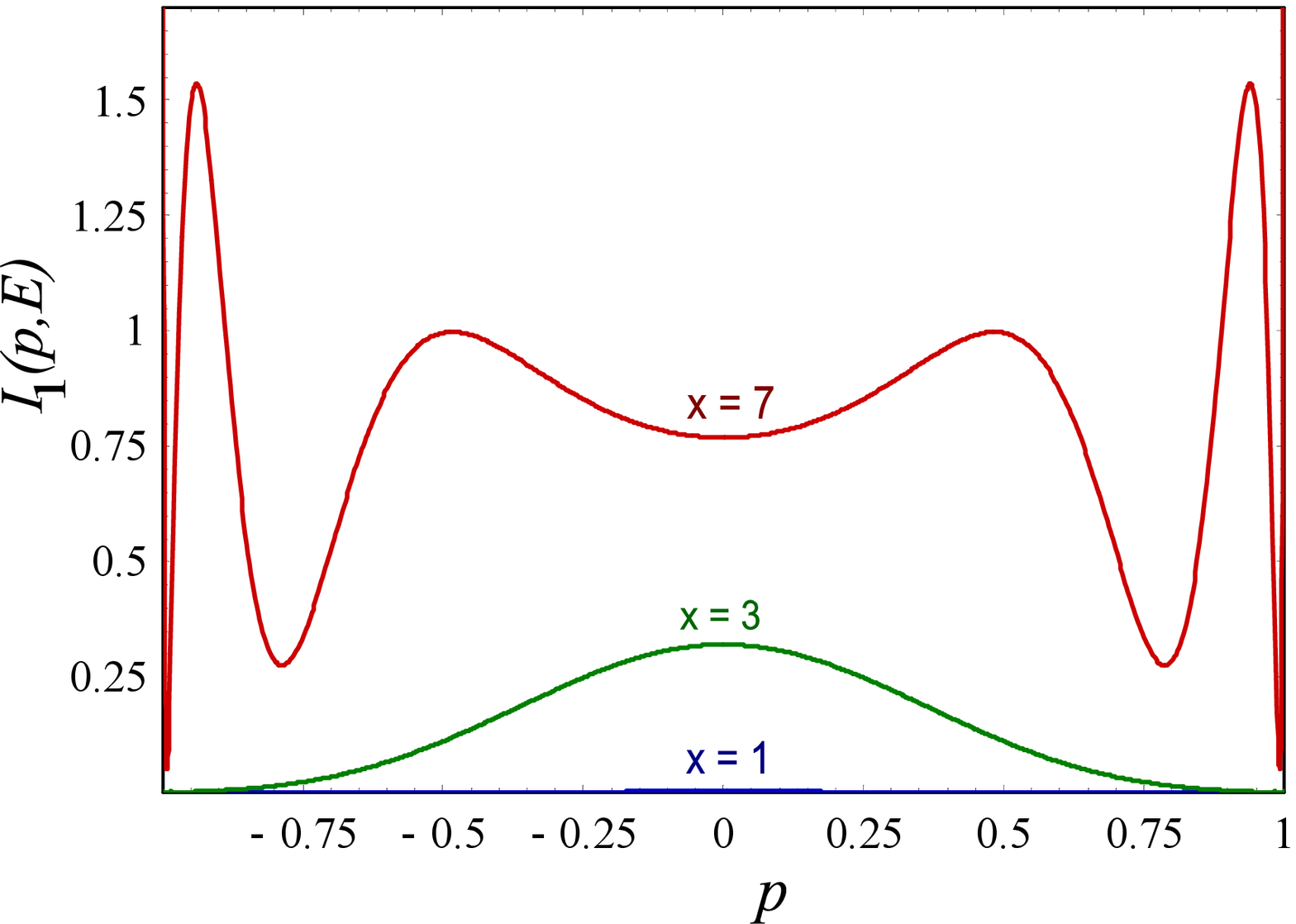}
\caption{The same as in Fig.2 but for $\mu_{\textrm{res}} D = 100$. }
\end{figure}
\begin{figure}[t]
\includegraphics[width=\linewidth]{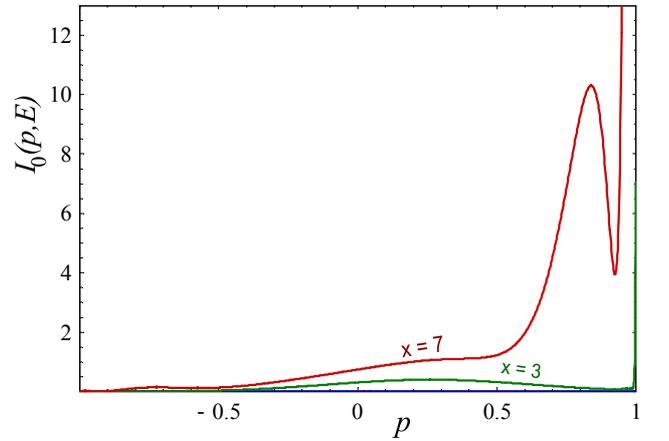}
\caption{The same as in Fig.3 but for $\mu_{\textrm{res}} D = 100$. }
\end{figure}

In order to illustrate the role of the resonant scattering we have
done numerical calculations for the symmetric Laue diffraction in an isotropic crystal
containing single resonant nucleus in every unit cell. We  neglected the potential
scattering amplitude compared to the resonant one. The latter  was described by Eq.~(\ref{eq:ap}) with the Debye-Waller
factor  $e^{-2W(Q)}=0.8$. In this approximation  the absorption
coefficient, depending on the detuning of the  resonance
$x=2(E-E_0)/\Gamma$, is written as
\begin{equation}
\mu(x)=\frac{\mu_{{\textrm{res}}}}{1+x^2},
\end{equation}
where $\mu_{{\textrm{res}}}$ is the resonant value of the
absorption coefficient, given by
\begin{equation}\label{}
\mu_{{\textrm{res}}}=\frac{4\pi}{\kappa^2v_0}\left(\frac{2I_e+1}{2I_g+1}\right)\frac{\Gamma_n}{\Gamma}.
\end{equation}

  As to the  functions $1/\tau_L$ and $1/\sigma_L$, they are determined by the following expressions:
\begin{eqnarray}
\frac{1}{\tau_L}=-\frac{\mu_{{\textrm{res}}}}{2\pi\gamma_0}\frac{xe^{-W(
Q)}}{1+x^2},\qquad \frac{1}{\sigma_L}=\frac{\mu_{{\textrm{res}}}}{2\pi\gamma_0}\frac{e^{-W(
Q)}}{1+x^2}.
\end{eqnarray}

Intensities of the diffracted and refracted beams as functions of $p$, calculated in units of $e^{-\mu_{\textrm{res}}D/\gamma_0}$ with $\gamma_0 \approx 1$
and $G_a=1 $ are presented in Figs.~ 2, 3 for $\mu_{\textrm{res}}D=20$ and in Figs.~4, 5 for more thick crystal with $\mu_{\textrm{res}}D=100$.

\section{Averaged intensities}
Remind that up to now we dealt with the waves ejected from the thread-like source with the coordinates $x=y=0$  and $z$ changing from $-\infty$ to $\infty$.
And now we shall analyze the role of  finite width $l$ of the entrance slit, regarding it as a sum of the parallel thread-like sources, spread over the interval $-l/2<y<l/2$, which corresponds to variation of the coordinate $p$ in the interval of the width $ \Delta p =l/L$. In the case of symmetric diffraction $\Delta p= l \cot\theta_{{\textrm{B}}}/D$.

The neutron wave in any point $p$ of the scanning slit is a superposition of the waves emitted by every such thread and afterwards passing the crystal region, confined by their own Borrmann triangle. The resulting waves in the point $p$ will be
\begin{equation}
\tilde{\Psi}^{(\nu)}_E(p)=\int_{-\Delta p/2}^{\Delta p/2}\Psi_E^{(\nu)}(p+\xi)d\xi,
\end{equation}
The corresponding integral intensity
\begin{equation}
\tilde{I}_{\nu}(p)=\int_0^{\infty}
|G_e(E)|^2|\tilde{\Psi}^{(\nu)}_E(p)|^2dE.
\end{equation}

 Besides, when the scanning slit  has the
width $l$, the intensity $\tilde{I}_{\nu}(p)$ should be
integrated over $p$ from $\bar{p}-\Delta p/2$ to $\bar{p}+\Delta p/2$, where $\bar{p}$ denotes the  coordinate of the slit's middle. So the flux of neutrons per unit time, emerging from the scanning slit
in the $\nu$th direction, is determined by
\begin{eqnarray}
J_{\nu}(\bar{p})=\bar{v}
\int_{\bar{p}-\Delta p/2}^{\bar{p}+ \Delta p/2}
\tilde{I}_{\nu}(p)dp,
\end{eqnarray}
where
$\bar{v}$  is the velocity  bound to the average energy $\bar{E}$:
\begin{equation}
\bar{v}=\hbar\bar{\kappa}/m, \qquad \bar{\kappa}=\sqrt{2m\bar{E}}/\hbar.
\end{equation}

We compared our results with  the data of Shull \cite{Shull1}, who had observed the symmetric Laue diffraction of  neutrons at (111) planes of silicon crystal. It has the diamond  structure
with the crystal constant $a=5.4311$~\AA \; and contains 8 atoms
in the elementary cell \cite{Si}. A spacing of the adjacent (111)
planes  equals $d=a/\sqrt{3}$.  The corresponding scattering
amplitudes are
\begin{eqnarray}
F({\boldsymbol\kappa}_0,{\boldsymbol\kappa}_0)=
F({\boldsymbol\kappa}_1,{\boldsymbol\kappa}_1)=
-8{\bar b},\\
 (F({\boldsymbol\kappa}_0,{\boldsymbol\kappa}_1)F({\boldsymbol\kappa}_1,{\boldsymbol\kappa}_0))^{1/2}=4\sqrt{2}\cdot{\bar b}e^{-W(Q)},
\end{eqnarray}
where the scattering vector $Q=h=2\pi n/d$  and $n$ is an integer. We took $n=1$.

In numerical calculations of the intensity
$ I_1(p;E)$ we used the following experimental parameters:   the wave length  $\lambda=1.034$~\AA, crystal thickness $D=0.3315$~cm,  the coherent scattering length $\bar{b}=0.41786\cdot 10^{-12}$~cm and the  factor $e^{-W(Q)}= 0.9984$ \cite{Shull1}.
The calculated dependence  of  $ I_1(p;E)$ on $p$  is drawn in Fig.~6, where it is clearly
seen a fringe structure, which becomes more and more dense as $p$ approaches margins of the Borrmann triangle.
These calculations have been performed in Kato's approximation  $\sigma_a>>|\Delta\vartheta|$, ignoring finite width of the slits.

The experimental curves \cite{Shull1} manifest weak oscillations only in the central part of the Borrmann triangle at $p$ close to zero.
In order to reproduce them we made calculations of the diffracted neutron flux
$J_1(\bar{p})$ with the above parameters and $l=0.13$ mm for both slits, taking the dispersions  $\sigma_e=1.8\cdot 10^{-3}$~eV and  $\sigma=0.0025$.
The theoretical curve, averaged in such manner, is
compared with the experimental data of Shull in Fig.~7. Note that all the experimental data are estimated very roughly by scanning the paper \cite{Shull1}.

\begin{figure}[t]
\includegraphics[width=\linewidth]{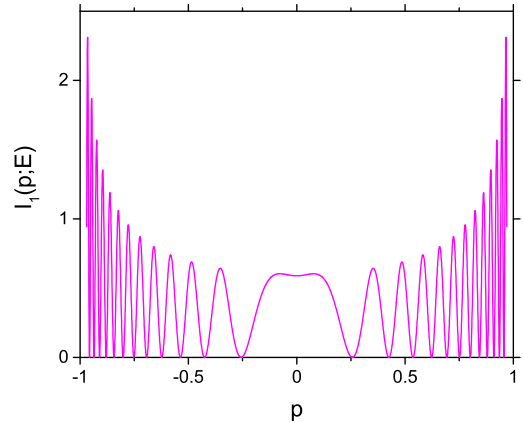}
\caption{The intensity of monochromatic neutron beam, diffracted at the (111) planes of silicon crystal, versus the reduced coordinate $p$.
The wave length of neutrons $\lambda=1.034$~\AA. }
\end{figure}

\begin{figure}[t]
\includegraphics[width=\linewidth]{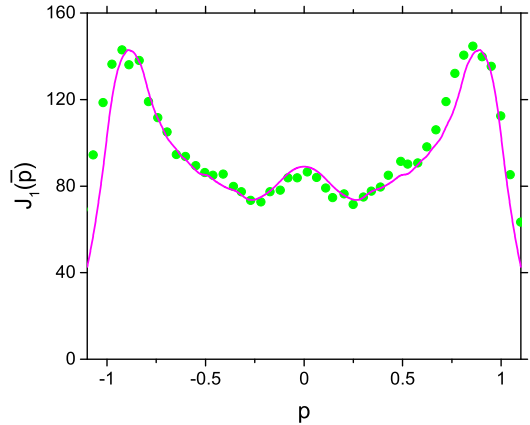}
\caption{The averaged flux (in arbitrary units)  of  neutrons, diffracted at the (111)  planes of silicon crystal, emerged from  the exit slit.
The experimental data \cite{Shull1} are indicated by circles, the calculations by solid line. }
\end{figure}

\section{Conclusion}
We have built general theory of the Laue diffraction  in perfect crystals of low-energy neutrons, emerging from the narrow entrance slit.
The  resonant neutron scattering by the nuclei is accounted in analogy with the theory \cite{Kagan}.
In addition, we included in our equations the angular distribution of incident neutrons $G_a(\theta)$ with arbitrary dispersion $\sigma$, which may be of the order of the diffraction angular interval $|\Delta\vartheta|$, whereas the Kato's theory [26-29] only treats the case $\sigma>>|\Delta\vartheta|$.
In this limiting case in the absence of resonances the derived formulas well correlate with those of the dynamical diffraction theory of x-rays in perfect crystals \cite{Authier}.

In Figs.~(2)--(5)  are shown the transmission and diffraction patterns, calculated in the vicinity of the isolated resonance for several values of the resonance detuning $x$. They resemble the pictures of the x-ray optics \cite{Authier}. Namely, far from the resonance, $x>>1$, as the crystal absorbs weakly, the intensity curve $I_1(p;E) $ of the diffracted beam tends to infinity at the edges of the Borrmann triangle, $ p \to \pm 1$, while the refracted beam is mainly concentrated in the forward direction, $p \to 1$. At the same time,  approaching the resonance, $x \to  0$,  as the absorbtion increases, both curves are mainly concentrated in the center of the Borrmann triangle $p=0$.
If there is strong nuclear absorption, of two waves passing the crystal in the forward or diffracted direction only weakly absorbed wave reaches the exit crystal surface. According to Eqs.~(\ref{eq:F11}), (\ref{eq:F12})  its intensity
\begin{equation}
I_{\nu}(p;E) \sim \exp\left(-\mu_s D+\frac{\pi D}{\sigma_L}\sqrt{1-p^2} \right),
\end{equation}
where $\nu=0$ or 1.
From here we see that this function has maximum at $p=0$ and  falls down with growing $|p|$.
Hence, suppression of the neutron capture by the nuclei weakens with deviation from the center of the Borrmann triangle.
In other words, the neutron waves move inside the crystal mainly along the reflecting planes.
This qualitatively explains the behavior of curves in Figs. (2)--(4) with small $x$, which describe diffraction close to  the resonance.

While in weakly absorbing crystal the interference beats over the back crystal surface appear very explicitly (see Fig.~6), they become smashed out in the case of strong absorption as it is shown in Figs.~2--4. Most explicit fringe structure appears at the wing of the resonance as  $x=7$ in thick crystal with $\mu_{\textrm{res}} D=100$ (see Figs.~4, 5). The curves for $x=0$ and 1  practically coincide there with the axis $p$. It is curious that in Fig.~2 the curves $x=3$ and $x=7$ are  rearranged in the central part of the Borrmann triangle. It is one more manifestation of the same two-wave interference, reflected in the squared sine  in Eq.~(\ref{eq:F12}).
 The averaging of the diffracted beam intensity $I_1(p;E)$ in weakly absorbing crystals over the slits lifts its infinite  grows at $p \to \pm 1$
 and shifts the maxima of $I_1(p;E)$ inside the Borrmann triangle. The same effect is ensured by the angular distribution $G_a(\theta)$ with dispersion $\sigma$ comparable with the diffraction interval $|\Delta\vartheta|$. In the limiting case of $\sigma<<|\Delta\vartheta|$ the curve $I_1(p;E)$  collapses to a narrow peak at $p=0$ like delta function.

The experiments similar to those of Shil'shtein et al. \cite{exp2} on suppression of the neutron capture are desirable  for understanding
 the role of angular divergence of the neutron beams. Note that the experimental data \cite{exp2} considerably deviate from predictions of the plane-wave theory \cite{Kagan}.
Anyway, we hope that our theory would be used   for precise determination of the nuclear scattering lengths and parameters of neutron resonances
in the diffraction experiments similar to Shull's studies. It can be also  helpful for planning any other neutron-optical experiments like [33-37].

% Use the following code if you wish to generate your bibliography with BibTeX;
% replace the string "pss_demo" below with the name(s) of
% the BibTeX data base(s) you want to use.
% The resulting bibliography-output (the content of the .bbl file)
% must be pasted back into this file before submission.
% Please also include your BibTeX data base file(s) in your submission
% so that we can re-run BibTeX if necessary.
%
%\bibliographystyle{pss}
%\bibliography{pss_demo}
%
% Replace the following example bibliography with your references
% before submission:

\end{document}